\documentclass[aps,prl,superscriptaddress,showpacs,floatfix,nofootinbib,notitlepage,twocolumn]{revtex4-1}
\usepackage{amsmath,graphicx,float,hyperref,csquotes,mdframed}

\def\bra{\langle}
\def\ket{\rangle}

\newcommand{\trento}{T$\mathrel{\protect\raisebox{-2.1pt}{R}}$ENTo}

\newcommand{\pbpb}{$^{208}$Pb+$^{208}$Pb}

\newcommand{\uuuu}{$^{238}$U+$^{238}$U}
\newcommand{\xexe}{$^{129}$Xe+$^{129}$Xe}
\newcommand{\ruru}{$^{96}$Ru+$^{96}$Ru}
\newcommand{\zrzr}{$^{96}$Zr+$^{96}$Zr}

\usepackage{tikz}
\usetikzlibrary{tikzmark}

\begin{document}

\title{Accessing the shape of atomic nuclei with relativistic collisions of isobars}

\author{Giuliano Giacalone}
\affiliation{Institut f\"ur Theoretische Physik, Universit\"at Heidelberg,
Philosophenweg 16, 69120 Heidelberg, Germany}

\author{Jiangyong Jia}
\affiliation{Department of Chemistry, Stony Brook University, Stony Brook, NY 11794, USA}
\affiliation{Physics Department, Brookhaven National Laboratory, Upton, NY 11976, USA}

\author{Vittorio Som\`a}
\affiliation{IRFU, CEA, Universit\'e Paris-Saclay, 91191 Gif-sur-Yvette, France}

\begin{abstract}
Nuclides sharing the same mass number (isobars) are observed ubiquitously along the stability line. While having nearly identical radii, stable isobars can differ in shape, and present in particular different quadrupole deformations. We show that even small differences in these deformations can be probed by relativistic nuclear collisions experiments, where they manifest as deviations from unity in the ratios of elliptic flow coefficients taken between isobaric systems. Collider experiments with isobars represent, thus, a unique means to obtain quantitative information about the geometric shape of atomic nuclei.
\end{abstract}

\maketitle

\paragraph{{\bf Introduction.}}A remarkable connection between low- and high-energy nuclear physics has been recently established in collider experiments conducted at the BNL Relativistic Heavy Ion Collider (RHIC) and at the CERN Large Hadron Collider (LHC) with the realization that the output of relativistic nuclear collisions is strongly affected by the deformation of the colliding ions.

The key observable driving this observation is elliptic flow, the quadrupole deformation (second Fourier harmonic) of the azimuthal distribution of hadrons detected in the final state of relativistic nuclear collisions~\cite{Heinz:2013th}:
\begin{equation}
    V_2 \propto \int_{\rm detector} f(\varphi) e^{i2\varphi},
\end{equation}
where $f(\varphi)$ is the distribution of azimuthal angles (in momentum space) collected in a collision event. 

\begin{figure}[b]
    \centering
    \includegraphics[width=.73\linewidth]{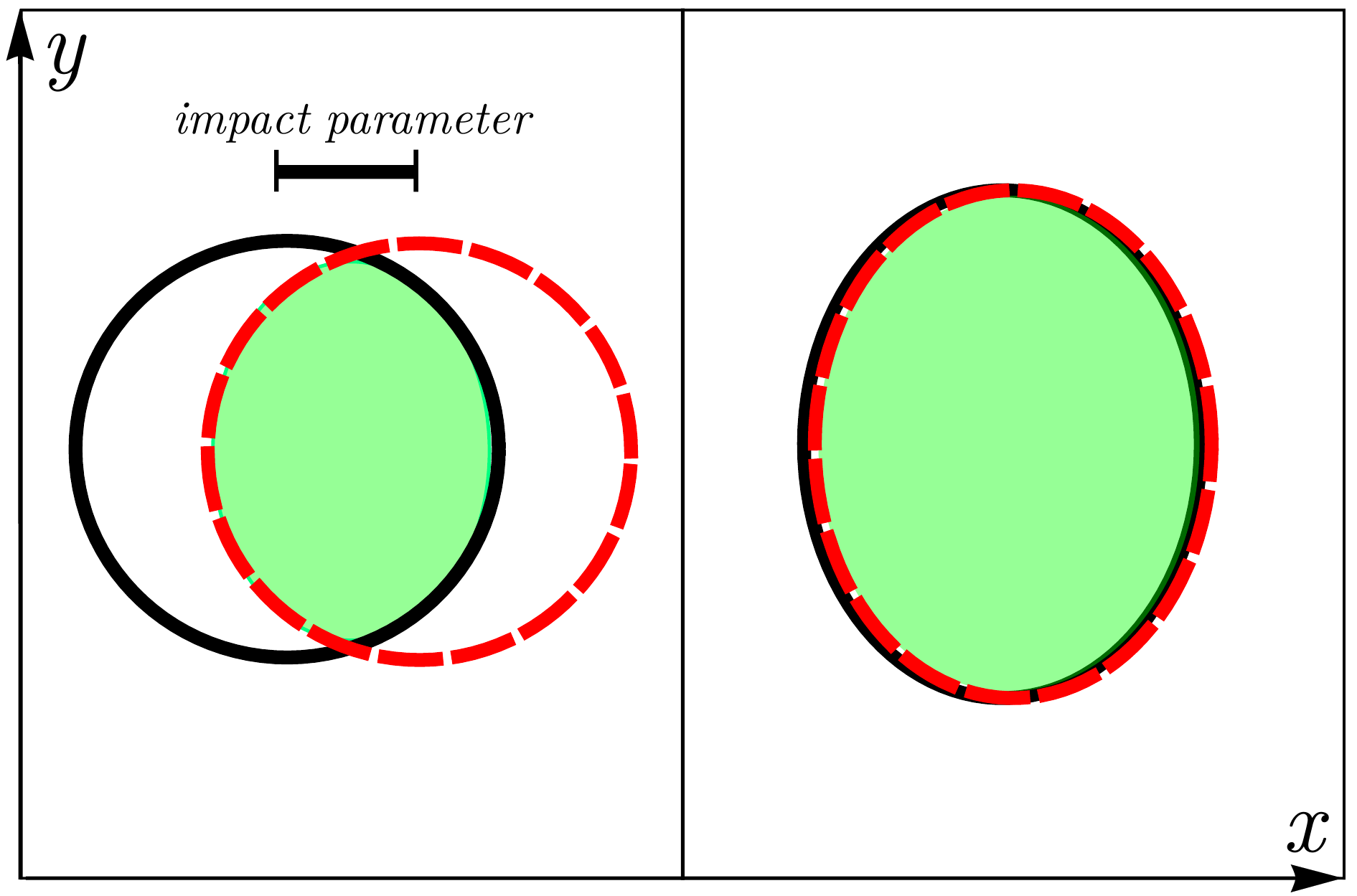}
    \caption{Anisotropic overlap regions in nuclear collisions. Left: a collision of spherical nuclei breaks anisotropy in the transverse plane due to the finite impact parameter. Right: a central collision of deformed nuclei breaks anisotropy due to the non-spherical shape of the colliding bodies.}
    \label{fig:1}
\end{figure}

In nucleus-nucleus collisions, elliptic flow emerges as a response to the quadrupole asymmetry (ellipticity) of the system created, right after the interaction takes place, in the plane transverse to the beam direction~\cite{Teaney:2010vd}:
\begin{equation}
    \mathcal{E}_2 \propto \int_{\rm overlap~area} \epsilon(r,\phi)~r^2 e^{i2\phi},
\end{equation}
where $(r,\phi)$ parametrize the transverse plane (for simplicity at $z=0$), and $\epsilon$ is the density of energy deposited in the overlap. In full generality, $\mathcal{E}_2 \neq 0 \Rightarrow V_2 \neq 0$. As illustrated in the left panel of Fig.~\ref{fig:1}, any collision occurring at finite impact parameter presents an overlap area which carries an elliptical deformation, i.e., $\mathcal{E}_2 \neq 0$, explaining in particular the observation that $V_2$ grows steeply with the collision impact parameter. 

However, for the majority of isotopes, even in the limit of vanishing impact parameter, one expects $\mathcal{E}_2 \neq 0$, and thus $V_2 \neq 0$ from nuclear structure arguments. Most of nuclei are in fact deformed objects, presenting a nonvanishing intrinsic quadrupole moment, i.e., an ellipsoidal deformation~\cite{BohrMottelson}:
\begin{equation}
    Q_{20} \propto \int_{\rm nucleus} \rho(r,\Theta,\Phi)~ r^2Y_{20}(\Theta,\Phi),
\end{equation}
where $\rho(r,\Theta,\Phi)$ represents the nucleon density in the intrinsic frame of the nucleus. Ultrarelativistic collisions take snapshots of randomly-oriented configurations of nucleons at the time of interaction, so that, if the colliding ions present $Q_{20}\neq 0$, regions of overlap such as that proposed in the right panel of Fig.~\ref{fig:1} can be produced. These lead to $\mathcal{E}_2\neq 0$ at zero impact parameter.

The bottom line is that in nucleus-nucleus collisions:
\begin{equation}
    Q_{20} \neq 0 ~~~\Longrightarrow~~~ \mathcal{E}_2 \neq 0 ~~~\Longrightarrow~~~  V_2 \neq 0.
\end{equation}
The importance of this statement has been recently clarified in the context of \uuuu{} collisions at RHIC, with the realization that observables based on $V_2$ (and on the hadron mean transverse momentum, $\bra p_t \ket$) are essentially dominated by effects due to the deformed shape of uranium nuclei~\cite{Adamczyk:2015obl,jia}, and at LHC with the measurement of an abnormally large $V_2$ in \xexe{} collisions compared to \pbpb{} collisions~\cite{Acharya:2018ihu,Sirunyan:2019wqp,Aad:2019xmh}.

These discoveries naturally trigger the question of whether one can use the great resolving power of high-energy colliders to infer something new about the low-energy structure of nuclei,
and provide a new means to test the state-of-the-art approaches applied to the nuclear many-body problem. In this Letter, we show that this is possible, and we delineate an experimental program to pursue this goal.

\paragraph{{\bf The idea.}}We exploit the seemingly uninteresting fact that a large number of stable nuclides belong to pairs of isobars, i.e., that for a given nuclide X one can often find a different nuclide Y that contains the same number of nucleons. This feature has an important implication for high-energy collisions. If X and Y are isobars, then X+X collisions produce a system which has the same properties (volume, density) as that produced in Y+Y collisions. As a consequence, X+X and Y+Y systems present the same geometry, the same dynamical evolution, and thus the same elliptic flow in the final state.

This leads us to our main point. Given two isobars, X and Y, we ask the following:
\begin{equation}
\boxed{
\label{eq:v2}
    \frac{v_2\{2\}_{{\rm X+X}} }{ v_2\{2\}_{\rm Y+Y} } \stackrel{?}{=} 1
}
\end{equation}
where $v_2\{2\}$ represents the usual rms measure of the magnitude of $V_2$ in a given multiplicity class. As argued above, the ratio should be equal to 1. Experimentally, once a number of minimum bias collisions of order $10^8$ is available, the ratio at small centralities can be obtained free of statistical error, while systematic uncertainties cancel in the ratio if the detector conditions of the X+X and Y+Y runs are the same~\cite{Adam:2019fbq}. Corrections to the ratio in Eq.~(\ref{eq:v2}) can further appear if the system produced in X+X and Y+Y collisions have different sizes, for instance, due to the fact that X and Y present different neutron numbers. However, two stable isobars can differ in matter radius by at most 0.5\%~\cite{Li:2019kkh}, leading to a negligible correction to Eq.~(\ref{eq:v2}), of order of few per mille.

Significant deviations ($>$1\%) from unity in the ratio of the $v_2$ coefficients are instead caused by the different deformations of the chosen isobars. Deformation reflects the collective organization of nucleons in the nuclear ground state and can quickly vary with proton and neutron numbers. In general, then, one does not expect two stable isobars to present the same deformed shape. The point we want to make in this Letter is the following:
\begin{displayquote}
\textit{
 Given two isobars, X and Y, if one measures $ \frac{v_2\{2\}_{{\rm X+X}} }{ v_2\{2\}_{\rm Y+Y} } > 1$, with a deviation from unity of order 1\% or larger, then one must conclude that nuclide X has a larger intrinsic quadrupole deformation than nuclide Y.
}
\end{displayquote}
This statement is based on two facts: $(i)$ that elliptic flow emerges from the elliptic anisotropy of the overlap area; $(ii)$ that nuclei in their ground states typically have nonvanishing intrinsic quadrupole moments. 
These are established features of nuclear physics that do not rely on any specific approximation or model.

Therefore, through measurements of the ratio in Eq.~(\ref{eq:v2}) one obtains an information about the relative deformation of the isobars. In the following, we show that this qualitative information can in fact be turned into a quantitative one, as even small differences in the quadrupole deformation of two isobars give rise to unambiguous and detectable effects in the ratio of the $v_2$ coefficients. As mentioned above, this observable is virtually devoid of experimental error and systematically accessible, under the same experimental conditions, throughout the Segr\`e chart.
Such features are hardly attainable in low-energy nuclear structure experiments. These measurements are thus expected to challenge the predictions of nuclear models tuned to low-energy experimental data in an unprecedented way.

\paragraph{{\bf Application.}} So far we have illustrated our idea through conceptual arguments. To get some intuition about the kind of results that will be obtained in collider experiments, we now perform quantitative calculations of Eq.~(\ref{eq:v2}) by choosing some models, namely, a standard parametrization for the deformed nuclear matter density, and a Glauber-type model for the collision process. 

A common parametrization of the nucleon density is the 2-parameter Fermi (2pF) distribution:
\begin{equation}
\label{eq:2pf}
    \rho(r,\Theta,\Phi) \propto  \frac{1}{1+\exp \left ( \left [r - R(\Theta,\Phi) \right] / a \right ) } ,
\end{equation}
where $a$ denotes the surface diffuseness and the half-density radius $R$ carries information about the deformed shape.
We characterize $R$ through a spherical harmonic expansion:
\begin{equation}
    R(\Theta,\Phi) =  R_0 \biggl [ 1+\beta Y_{2,0}(\Theta,\Phi) \biggr ],
\end{equation}
truncated, for the present application, at the axial quadrupole, $Y_{2,0}$. We shall neglect, then, potential corrections due to triaxial, $Y_{2,\pm 2}$, and hexadecapole, $Y_{4,0}$, deformations, which can be systematically added in future. We stress, however, that for well-deformed nuclei, the most important for our analysis, such corrections are subleading, and will not alter our conclusions. The coefficient $\beta$ quantifies\footnote{We assume here that nuclei have a fixed quadrupole deformation, while, in reality, shape fluctuations are normally present (to a small or large extent, depending on the ``softness'' of the nuclear system, see e.g. Ref.~\cite{Poves2020}).
While the qualitative statements discussed in the previous section are independent of this feature, we have checked that the quantitative results presented in this section do not change significantly if a distribution in $\beta$ is considered instead of the fixed value stated below~\cite{Bally2021}.} the ellipsoidal shape of the nucleus:
\begin{equation}
    \beta \simeq  \frac{4\pi}{5} \frac{ \int \rho(r,\Theta,\Phi)~ r^2Y_{20}(\Theta,\Phi) }{ \int \rho(r,\Theta,\Phi)~r^2  }.  
\end{equation}
Well-deformed nuclei, such as $^{238}$U, or stable nuclides with $150<A<180$, are characterized by $\beta \approx 0.3$.

Two nuclei, described as randomly oriented, deformed batches of nucleons sampled independently according to the 2pF distribution given in Eq.~(\ref{eq:2pf}), are then collided at ultrarelativistic energy. 
On a collision-by-collision basis, the energy density deposited in the process possesses a nonvanishing eccentricity, $\varepsilon_2\equiv |\mathcal{E}_2|$, which triggers the development of elliptic flow during the expansion of the system, resulting in the observed momentum anisotropy, $V_2$. Dubbing $v_2\equiv|V_2|$, at a given multiplicity (centrality) one has: $v_2 = \kappa_2 \varepsilon_2$, where $\kappa_2$ is a real coefficient that depends on the properties of the system (e.g., equation of sate and viscosity in a hydrodynamic model~\cite{Heinz:2013th}). Now, as anticipated, isobaric systems share the same physical properties, so that a crucial simplification occurs: $\kappa_2 [{\rm X+X}] = \kappa_2 [{\rm Y+Y}]$. As soon as we take a ratio between $v_2$ coefficients calculated in two different isobaric systems, then, the response factor $\kappa_2$ drops out. This in turn implies that:
\begin{equation}
\label{eq:eps2ratio1}
    \frac{v_2\{2\}_{{\rm X+X}} }{ v_2\{2\}_{\rm Y+Y} } = \frac{\varepsilon_2\{2\}_{{\rm X+X}} }{ \varepsilon_2\{2\}_{\rm Y+Y}} \stackrel{?}{=} 1 .
\end{equation}
The question of whether or not the measured ratio of $v_2$ coefficients is equal to unity boils down to whether the two isobaric system possess the same fluctuations of $\varepsilon_2$.

This reformulation of our question in terms of $\varepsilon_2$ fluctuations allows us now to employ a collision model to calculate Eq.~(\ref{eq:eps2ratio1}), and thus to perform a quantitative evaluation of the ratio of the $v_2$ coefficients to be measured at colliders. To do so, we use the default \trento{} model~\cite{Moreland:2014oya}, which has proven able to capture with good accuracy the effects of the quadrupole deformation of nuclei on elliptic flow data collected in \uuuu{} collisions. We perform this analysis for two pairs of isobars:
\begin{itemize}
    \item A pair of well-deformed rare-earth nuclei with nearly identical deformations, namely, $^{154}$Sm and $^{154}$Gd. For the 2pF density profile, we assume that the matter distribution is well described with parameters extracted from the measured charge density, a good approximation for stable nuclides. For both nuclei we employ $R=5.975$~fm and $a=0.59$~fm, motivated by the results of Ref.~\cite{DeJager:1987qc}. For the deformation parameters, we adopt values inferred from the measured transition probabilities of the electric quadrupole operator from the ground state to the first 2$^+$ state, tabulated, e.g., in Ref.~\cite{nndc}. One finds\footnote{Note that the definition of these ``experimental'' $\beta$ relies on model approximations (e.g., a sharp nuclear surface) that are not completely consistent with the use of Eq.~(\ref{eq:2pf}). We neglect here this possible mismatch, which is typically of order 5-10\% (see Ref.~\cite{Shou:2014eya} for more details).} $\beta=0.34$ for $^{154}$Sm and $\beta=0.31$ for $^{154}$Gd.
    \item A pair of lighter nuclei, $^{96}$Zr and $^{96}$Ru. These species are of particular relevance because a run of both \zrzr{} and \ruru{} collisions has been performed at RHIC in 2018~\cite{Adam:2019fbq}, and experimental results will be released shortly. For the 2pF of these nuclei we set $R=5.06$~fm for $^{96}$Zr and $R=5.03$~fm for $^{96}$Ru, taking into account the fact that $^{96}$Zr has an excess of 4 neutrons, while $a=0.52$~fm for both. The deformation parameters are instead $\beta=0.15$ for $^{96}$Ru, and $\beta=0.06$ for $^{96}$Zr~\cite{nndc}.
\end{itemize}

\begin{figure}[t]
    \centering
    \includegraphics[width=.95\linewidth]{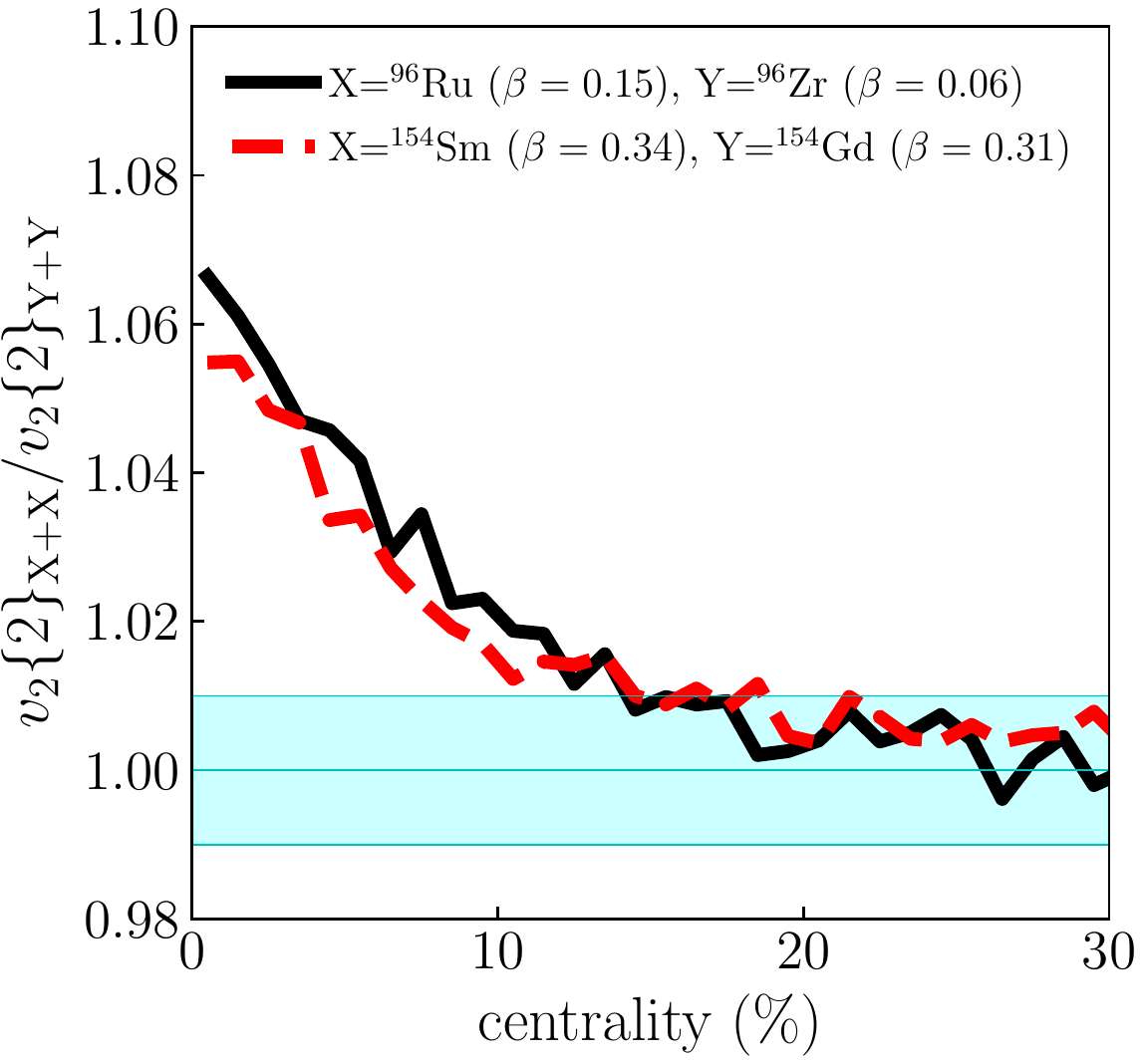}
    \caption{Rms elliptic flow in X+X collisions divided by the rms elliptic flow in Y+Y collisions as a function of collision centrality. The ratio of flow coefficients is estimated following Eq.~(\ref{eq:eps2ratio1}) and the \trento{} model. The shaded band represents a 1\% deviation from unity. Any deviation from unity which falls outside the shaded band can be considered as a significant signature that $\beta_{\rm X}\neq \beta_{\rm Y}$.}
    \label{fig:2}
\end{figure}

For each system we perform 5 million minimum bias collisions. We sort events into centrality classes according to the entropy created in the process, following the default \trento{} model prescription. In each centrality bin, we evaluate $\varepsilon_2\{2\}$, and by subsequently taking ratios between isobaric systems, as in Eq.~(\ref{eq:eps2ratio1}), we obtain the results displayed in Fig.~\ref{fig:2}. 

The shaded band corresponds to a departure from unity smaller than 1\%. Any deviation falling outside the band can be interpreted as a genuine signature of the different quadrupole deformations carried by the two isobars, although we caution that for precise comparisons with future experimental data, potential corrections due to the imperfect centrality definition of our model will have to be carefully addressed~\cite{Jia:2020tvb,Aaboud:2019sma}. The black solid line represents our result for the systems collided in 2018 at RHIC, and can be confronted with upcoming data. For our choice of the deformation parameters, we observe that the splitting between the flow coefficients in central collisions is well above 1\%, consistent with the fact that $^{96}$Ru has a larger quadrupole deformation in these simulations.  The same behavior is observed for the pair of heavier nuclei (red dashed line). A deviation from unity of order 5\% emerges in central collisions, due to the fact that $^{154}$Sm nuclei are more deformed. 

The fact that both pairs return a similar (small) splitting between $v_2$ coefficients can be understood as follows. At a given small centrality, one expects~\cite{Giacalone:2018apa}:
\begin{equation}
\label{eq:a0a1}
    \varepsilon_2\{2\}^2 = a_0 + a_1\beta^2.
\end{equation}
The coefficient $a_0$ is the eccentricity due to quantum fluctuations (nucleon positions), while the term proportional to $\beta^2$ represents the contribution from fluctuations of the geometry of the system driven by the random orientation of the deformed ions. Systems X+X and Y+Y have the same size and number of participant nucleons, therefore, they present the same coefficients $a_0$ and $a_1$. Now, in the \trento{} results we find that, even for $A=154$ and $\beta\approx0.3$, the contribution from $a_0$ in Eq~(\ref{eq:a0a1}) is larger than the contribution from $a_1\beta^2$. Inserting Eq.~(\ref{eq:a0a1}) in Eq.~(\ref{eq:eps2ratio1}), expanding the ratio around unity, and keeping the leading correction, equation~(\ref{eq:eps2ratio1}) becomes:
\begin{equation}
\label{eq:eps2ratio2}
    \frac{\varepsilon_2\{2\}_{{\rm X+X}} }{ \varepsilon_2\{2\}_{\rm Y+Y}} \simeq 1 + c \bigl ( \beta_{\rm X}^2 - \beta_{\rm Y}^2 \bigr ),
\end{equation}
where $c\sim\mathcal{O}(1)$. The order of the deviation from unity is driven by the difference $\beta_X^2 - \beta_Y^2$. For both pairs of isobars considered in our application one has $\beta_X^2 - \beta_Y^2\approx0.02$ (and $c\approx 3$), explaining the similarity between the curves shown in Fig.~\ref{fig:2}. 

Our most important result concerns the heavier species with $A=154$. The contribution from the nuclear deformation in Eq.~(\ref{eq:a0a1}) is quadratic in $\beta$, therefore, it is much more important for well-deformed nuclei, $\beta\approx0.3$. For this reason, and as demonstrated by the results in Fig.~\ref{fig:2}, for well-deformed nuclei even differences at the level of few percents in the values of $\beta$ will leave visible signatures in the ratio of flow coefficients. As anticipated, then, in this scenario the qualitative statement that X is more deformed than Y turns into a nontrivial quantitative issue, driven by small differences in the shape of the isobars. 

\paragraph{{\bf Conclusion \& Outlook.}}Thanks to the great imaging power of high-energy colliders, relativistic collision experiments involving stable isobars, such as \zrzr{} and \ruru{} collisions recently run at RHIC, yield ratios of $v_2$ coefficients between isobaric systems that are not simply equal to one, but rather look like the curves shown in Fig.~\ref{fig:2}. A given ratio falling outside the shaded band indicates that the geometric shapes of the colliding ions are different. This information is virtually free of experimental error, and has to be confronted with our knowledge of nuclear physics across energy scales.

Within the present paradigm, deviations from unity in the proposed ratio inform us about the relative quadrupole deformation of the colliding species. In the regime of well-deformed nuclei, the mere statement that nucleus X is more deformed than nucleus Y translates into precise, quantitative information, which can be accessed systematically across the nuclide chart thanks to the abundance of stable isobars found in Nature.

\begin{table}[t]
\centering
\begin{tabular}{|c|c|c|c|c|c|c|}
\hline
$A$ & isobars & $A$ & isobars & $A$ & isobars \cr
\hline
36 & Ar, S &  106 & Pd, Cd  & \tikzmark{148} 148 & Nd, Sm  \\
40 & Ca, Ar  & 108  & Pd, Cd & \tikzmark{150} 150  & Nd, Sm \tikzmark{Sm2} \\
46  & Ca, Ti  & 110  &  Pd, Cd  & 152 & Sm, Gd \\
48   & Ca, Ti  & 112  & Cd, Sn  & 154 & Sm, Gd \\
50 & Ti, V, Cr  & 113 & Cd, In  & 156  &  Gd, Dy \\
54 & Cr, Fe  & 114 & Cd, Sn  & 158  & Gd, Dy \\
64  & Ni, Zn  & 115 & In, Sn  & 160  & Gd, Dy  \\
70  & Zn, Ge  &  116 & Cd, Sn  & 162  & Dy, Er  \\
74  &  Ge, Se &  120 & Sn, Te  & 164  & Dy, Er \\
76  & Ge, Se  &  122 & Sn, Te  & 168 & Er, Yb  \\
78  &  Se, Kr &  123 & Sb, Te  & 170  & Er, Yb  \\
80  & Se, Kr  &  124 &  Sn, Te, Xe & 174  & Yb, Hf  \\
84  & Kr, Sr, Mo  &  126 & Te, Xe  & 176 & ~Yb, Lu, Hf~ \\
86  & Kr, Sr  & 128  & Te, Xe  & 180  & Hf, W \\
87  & Rb, Sr  & 130 & Te, Xe, Ba & 184  & W, Os \tikzmark{Os} \\
92  &  Zr, Nb, Mo & 132  & Xe, Ba  & 186 & W, Os \\
94  & Zr, Mo  & 134  & Xe, Ba & 187 & Re, Os \\
96  &  ~Zr, Mo, Ru~  & 136  & ~Xe, Ba, Ce~  & 190 & Os, Pt \\
98  & Mo, Ru  &  138 & Ba, La, Ce & 192  & Os, Pt \\
100  & Mo, Ru  & 142  & Ce, Nd  & 198 & Pt, Hg \\
102  & Ru, Pd  & 144  & Nd, Sm  & 204 & Hg, Pb \\
104  & Ru, Pd  & \tikzmark{146} 146  & Nd, Sm \tikzmark{Sm}  &  & \\
\hline
\end{tabular}
\begin{tikzpicture}[overlay,remember picture]
  \draw[line width=0.25mm,red] ([shift={(-.6ex,2.5ex)}]pic cs:150) rectangle ([shift={(3.2ex,-1ex)}]pic cs:Os);
\end{tikzpicture}
\begin{tikzpicture}[overlay,remember picture]
  \draw[line width=0.25mm,blue] ([shift={(-.6ex,2.5ex)}]pic cs:146) rectangle ([shift={(2.8ex,-.9ex)}]pic cs:Sm);
\end{tikzpicture}
\begin{tikzpicture}[overlay,remember picture]
  \draw[line width=0.25mm,blue] ([shift={(-.6ex,2.5ex)}]pic cs:148) rectangle ([shift={(2.7ex,-1ex)}]pic cs:Sm2);
\end{tikzpicture}
\caption{\label{tab:1} 
Pairs and triplets of stable isobars (half-life longer than $10^8~y$). A total of 139 nuclides are listed. The region highlighted in red contains large well-deformed nuclei ($A\geq150$, $\beta>0.2$). The region highlighted in blue corresponds instead to nuclides which may present as well an octupole deformation in their ground state.}
\end{table}

Physicists should take advantage of this great opportunity. All the pairs and triplets of isobars which are stable enough to be used in potential future collider experiments are listed in Tab.~\ref{tab:1}. Well-deformed nuclei are highlighted in a red box in the table. A recent study~\cite{Cao:2020rgr} further suggests that $^{146,148,150}$Nd and $^{150}$Sm may present an octupole deformation ($Q_{30}\propto \int r^3 Y_{30}(\Theta,\Phi) \rho(r,\Theta,\Phi) \neq 0$) in their ground state. A small octupole deformation would be visible in high-energy collisions as an enhancement of the fluctuations of triangular flow, $v_3$, as explicitly shown in a recent application to \pbpb{} collisions~\cite{Carzon:2020xwp}. Nd and Sm isotopes are therefore ideal candidates for such a study.

We stress that these experiments can be repeated for several pairs of isotopes in identical conditions and provide us with information that does not rely on specific nuclear structure details.
Ultrarelativistic collisions thus represent an outstanding tool that is truly complementary to modern low-energy experiments. They offer a unique way to test the goodness of existing nuclear models for a wide range of species, and consequently pose a solid baseline for the next generation of theory-to-data comparisons including \textit{ab-initio} frameworks of nuclear structure currently under intense development~\cite{Hergert:2020bxy}.

\paragraph{{\bf Acknowledgments.}} We would like to thank the participants of the Initial Stages 2021 conference, in particular Jaki Noronha-Hostler, S\"oren Schlichting, and Peter Steinberg for their feedback and inspiring comments on the topic of this manuscript. G.G. acknowledges useful discussions with Benjamin Bally, Michael Bender, and Matt Luzum. The work of G.G. is supported by the Deutsche Forschungsgemeinschaft (DFG, German Research Foundation) under
Germany’s Excellence Strategy EXC 2181/1 - 390900948
(the Heidelberg STRUCTURES Excellence Cluster),
SFB 1225 (ISOQUANT) and FL 736/3-1. The work of J.J is supported by DOE DEFG0287ER40331 and NSF PHY-1913138.

\end{document}